\documentclass[a4paper,11pt]{article}
\usepackage{pos}
\usepackage{amsmath}
\usepackage{placeins}
\usepackage{multirow}
\usepackage{tabularx}

\title{Existence and Non-Existence of Doubly Heavy Tetraquark Bound States}

\author*[a]{Martin Pflaumer}
\author[b,c]{Luka Leskovec}
\author[d]{Stefan Meinel}
\author[a,e]{Marc Wagner}

\affiliation[a]{Institut f\"ur Theoretische Physik, Goethe-Universit\"at Frankfurt am Main, \\ Max-von-Laue-Stra{\ss}e 1, D-60438 Frankfurt am Main, Germany}

\affiliation[b]{Thomas Jefferson National Accelerator Facility,\\
	 Newport News, VA 23606, USA}

\affiliation[c]{Department of Physics, Old Dominion University,\\
	 Norfolk, VA 23529, USA}

\affiliation[d]{Department of Physics, University of Arizona,\\
	 Tucson, AZ 85721, USA}

\affiliation[e]{Helmholtz Research Academy Hesse for FAIR,\\
	 Campus Riedberg, Max-von-Laue-Stra{\ss}e 12, D-60438 Frankfurt am Main, Germany}

\emailAdd{pflaumer@itp.uni-frankfurt.de}
\emailAdd{leskovec@jlab.org}
\emailAdd{smeinel@arizona.edu}
\emailAdd{mwagner@itp.uni-frankfurt.de}

\abstract{In this work we investigate the existence of bound states for doubly heavy tetraquark systems $ \bar{Q}\bar{Q}'qq' $ in a full lattice-QCD computation, where heavy bottom quarks are treated in the framework of non-relativistic QCD. We focus on three systems with quark content $ \bar{b}\bar{b}ud $, $ \bar{b}\bar{b}us $ and $ \bar{b}\bar{c}ud $. We show evidence for the existence of $ \bar{b}\bar{b}ud $ and $ \bar{b}\bar{b}us $ bound states, while no binding appears to be present for $ \bar{b}\bar{c}ud $. For the bound four-quark states we also discuss the importance of various creation operators and give an estimate of the meson-meson and diquark-antidiquark percentages.}

\FullConference{%
 The 38th International Symposium on Lattice Field Theory, LATTICE2021
  26th-30th July, 2021
  Zoom/Gather@Massachusetts Institute of Technology
}

\newcommand{\op}{\mathcal{O}}
\newcolumntype{C}[1]{>{\centering\arraybackslash}m{#1}}

\begin{document}
\maketitle

%
\section{Introduction}
In the last decade, several hadrons that cannot be described by an ordinary quark-antiquark pair were observed in experiments. Their quantum numbers are, however, consistent with a four-quark structure. One prominent example is the electrically charged states $Z_b(10610)^+$ and $Z_b(10650)^+$ \cite{Belle:2011aa}. Their masses and decay channels strongly suggest a $\bar b b$ pair, but their non-vanishing electrical charge indicates the presence of another light quark-antiquark pair.

These experimental results triggered many theoretical studies of tetraquarks, which are often extremely challenging, in particular when several decay channels exist. In this work we are focusing on less difficult four-quark systems composed of two heavy antiquarks $\bar{Q}\bar{Q}'$ with $Q,Q' \in \{b,c\}$ and two light quarks $qq'$ with $q,q' \in \{u,d,s\}$. This particular quark structure $\bar{Q}\bar{Q}'qq'$ is very promising with respect to the formation of hadronically stable tetraquarks, as there is evidence that in the limit of large heavy quark masses such tetraquarks exist (see e.g.\ Refs.\ \cite{Carlson:1987hh,Manohar:1992nd,Eichten:2017ffp,Karliner:2017qjm}).

In previous lattice-QCD studies the Born-Oppenheimer approximation was used extensively to investigate the $\bar{b}\bar{b}ud$ system. Those studies predicted a bound state with quantum numbers $I(J^P)=0(1^+)$ and binding energy $\approx 60 \, \textrm{MeV} \ldots 90 \, \textrm{MeV}$ \cite{Brown:2012tm,Bicudo:2012qt,Bicudo:2015kna,Bicudo:2015vta,Bicudo:2016ooe,Bicudo:2021qxj}. Moreover, a resonance with quantum numbers $I(J^P)=0(1^-)$ was found, which has a resonance energy $\approx 20 \, \textrm{MeV}$ above the $B B$ threshold and a width $\Gamma \approx 100 \, \textrm{MeV}$ \cite{Bicudo:2017szl}. More rigorous full lattice-QCD studies recently confirmed the hadronically stable $\bar{b}\bar{b}ud$ tetraquark and predicted another bound state for $\bar{b}\bar{b}us$, while the situation is less clear for $\bar{b}\bar{c}ud$ \cite{Francis:2016hui,Francis:2017bjr,Francis:2018jyb,Junnarkar:2018twb,Leskovec:2019ioa,Hudspith:2020tdf,Pflaumer:2020ogv}. In the following we give an update on our ongoing full lattice-QCD investigations of $\bar{b}\bar{b}ud$, $\bar{b}\bar{b}us$ and $\bar{b}\bar{c}ud$ tetraquarks.
%
%
\section{Lattice Setup}
We use gauge-link configurations generated by the RBC and UKQCD collaborations with $2+1$ flavors of domain-wall fermions and the Iwasaki gauge action \cite{Aoki:2010dy,Blum:2014tka}. Details of the ensembles are collected in Tab.\ \ref{tab:configurations}. They differ in the lattice spacing, the lattice extent and the pion mass. One of the ensembles has a pion mass equal to the physical pion mass. In the following we show and discuss results only for selected ensembles, but computations were always performed on all five ensembles, e.g.\ to study the pion-mass dependence of the binding energy and finite volume effects via a scattering analysis for the $\bar{b}\bar{b}ud$ system (see Ref.\ \cite{Leskovec:2019ioa} for details).

\begin{table}[htb]
	\centering
	\begin{tabular}{cccccc} \hline \hline 
		Ensemble & $N_s^3 \times N_t$ & $a$ [fm] 	& $a m_{u;d}$ & $a m_{s}$ & $m_\pi$ [MeV]  \\ \hline
		C00078& $48^3 \times 96$ & $0.1141(3)$	& $0.00078$			  & $0.0362$           & $139(1)$ \\ \hline
		C005 & $24^3 \times 64$	 & $0.1106(3)$	& $0.005\phantom{00}$ & $0.04\phantom{00}$ & $340(1)$ \\
		C01	 & $24^3 \times 64$	 & $0.1106(3)$  & $0.01\phantom{000}$ & $0.04\phantom{00}$ & $431(1)$ \\ \hline
		F004 & $32^3 \times 64$  & $0.0828(3)$  & $0.004\phantom{00}$ & $0.03\phantom{00}$ & $303(1)$ \\
		F006 & $32^3 \times 64$  & $0.0828(3)$	& $0.006\phantom{00}$ & $0.03\phantom{00}$ & $360(1)$ \\ \hline\hline
	\end{tabular}
	\caption{\label{tab:configurations}Gauge-link ensembles \cite{Aoki:2010dy, Blum:2014tka} used in this work. $N_s$, $N_t$: number of lattice sites in spatial and temporal directions; $a$: lattice spacing; $am_{u;d}$: bare up and down quark mass; $a m_{s}$: bare strange quark mass; $m_\pi$: pion mass. }
\end{table}

We used spatially smeared point-to-all propagators for all quark flavors. Bottom propagators are computed in the NRQCD framework \cite{Thacker:1990bm,Lepage:1992tx} and charm propagators correspond to a relativistic heavy quark action \cite{Brown:2014ena}.
%
%
\section{Interpolating Operators}
Two distinct types of interpolating operators are used in our investigation. The first type corresponds to local operators, where all four-quarks are centered at the same point in space. We consider local meson-meson as well as local diquark-antidiquark structures. The second type of interpolating operators corresponds to non-local or scattering operators. They describe two spatially separated independent mesons. In case the ground state in a given sector is a four-quark bound state, we expect that the local operators will generate a good overlap to that state. Since meson-meson scattering states are expected to be rather close, we consider it extremely important to also include scattering operators. Only the combination of both types of interpolating operators might allow to accurately resolve all low lying states and to isolate a possibly existing stable tetraquark from scattering states.

In detail, our interpolating operators are given by
\begin{alignat}{2}
	&\op_{\textrm{loc,}MM} \propto && \sum_{\mathbf{x}} \, M_1(\mathbf{x}) \,M_2(\mathbf{x}) \\
	&\op_{\textrm{loc,}Dd} \propto &&\sum_{\mathbf{x}} \, \bar{Q}^a_1 \gamma_j \mathcal{C} \bar{Q}^b_2(\mathbf{x}) \, q^a_1 \mathcal{C} \gamma_5 \Gamma_2 q^b_2(\mathbf{x}) \label{EQN_opDd} \\
	&\op_{\textrm{scatt,}MM} \propto && \sum_{\mathbf{x}} \, M_1(\mathbf{x}) \, \sum_{\mathbf{y}}\,M_2(\mathbf{y}),
\end{alignat}
where we use the notation $ M_j(\mathbf{x}) =  \bar{Q}_j \Gamma_j q_j(\mathbf{x})$ for mesonic interpolators and $\mathcal{C}$ denotes the charge conjugation matrix. For each flavor and $I(J^P)$ sector, we consider the diquark-antidiquark operator \eqref{EQN_opDd} and several meson-meson operators as listed in Tab.\ \ref{tab:mesonic_operators}. For $\bar{b}\bar{b}ud$ and $\bar{b}\bar{c}ud$ we study $I = 0$. The anti-symmetric flavor combination is realized via $\sum_{\mathbf{x}} \, (M_1(\mathbf{x}) \,M_2(\mathbf{x}) - u \leftrightarrow d)$.

\begin{table}[h]
	\centering
	\begin{tabular}{C{1.5cm}C{2.5cm}C{3cm}C{2.5cm}C{2.5cm}}
	 				& $ \bar{b}\bar{b}ud $ 	& $ \bar{b}\bar{b}us $ 	& \multicolumn{2}{c}{$ \bar{b}\bar{c}ud $}  \\ \hline \hline 
	   $ I(J^P) $ 	& $ 0(1^+) $ 			& $ \tfrac{1}{2}(1^+) $ & $ 0(0^+) $ 		& $ 0(1^+) $ \\ \hline 
	    		 	& $ BB^\ast $, $ B^\ast B^\ast $
	   				& $ B_sB^\ast $, $ B_s^\ast B $, $ B_s^\ast B^\ast $
	   				& $ BD $
	   				& $ BD^\ast $, $ B^\ast D $ \\ \hline \hline 
	\end{tabular}
\caption{\label{tab:mesonic_operators}Meson-meson operators considered for each flavor and $I(J^P)$ sector. For pseudoscalar mesons we use $\Gamma_{1,2} = \gamma_5$, for vector mesons we use $\Gamma_{1,2} = \gamma_j$.}
\end{table}

%
\section{Energy Levels of the $\bar{Q}\bar{Q'}qq'$ Systems}
\begin{figure}[htb]
	\centering
	\includegraphics[width=0.65\textwidth, trim = 15 15 15 15 ,clip]{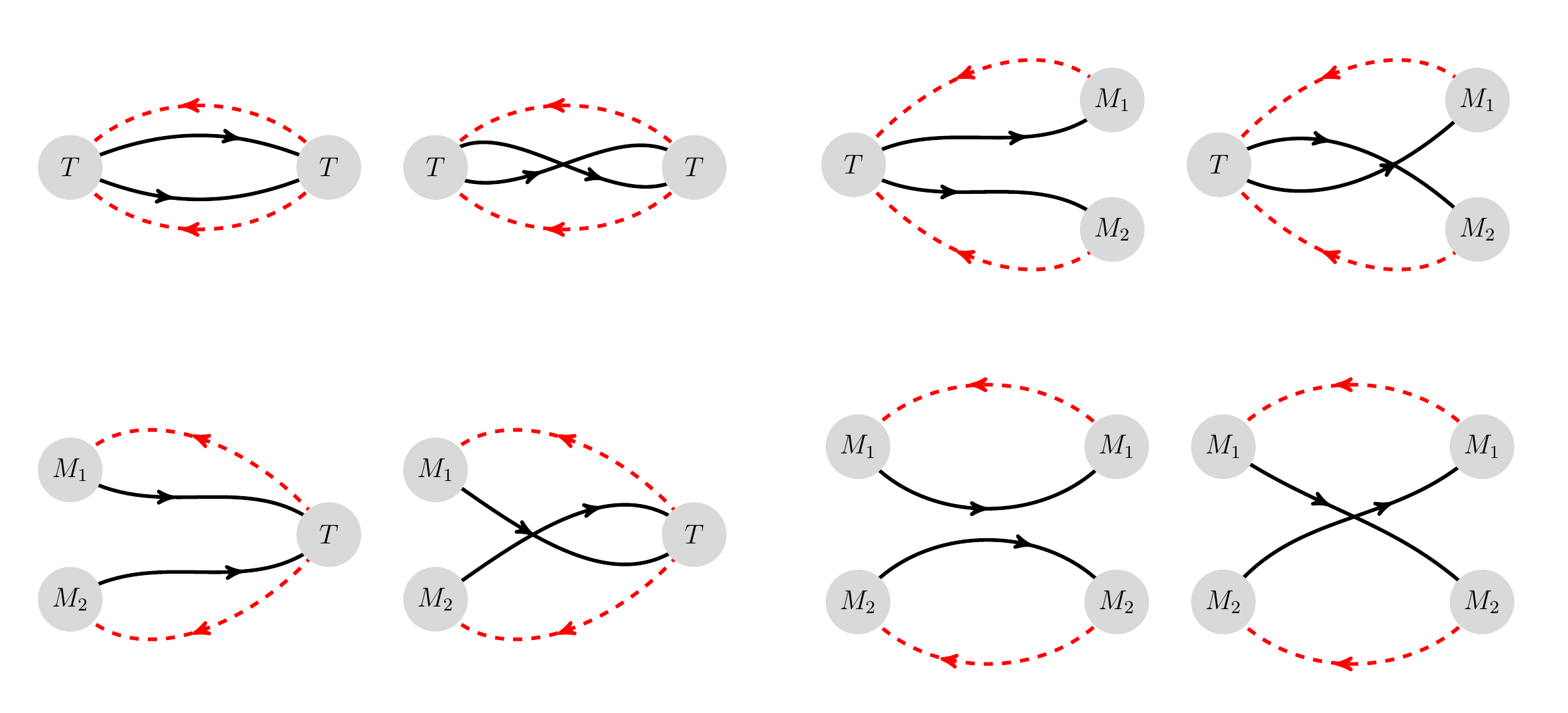}
	\caption{\label{fig:Wick_contractions}Schematic representation of Wick contractions for different types of correlation matrix elements. $T$ represents a local tetraquark operator and $M_1$ and $M_2$ represent the two mesons forming a scattering operator. Black lines correspond to heavy quark propagators, red lines to light quark propagators.}
\end{figure}

We computed correlation matrices $C_{jk}(t) = \langle  \op_j(t) \op_k^\dagger(0)\rangle$, where $\op_j$ and  $\op_k$ are the interpolating operators defined in the previous section. We present a schematic representation of the necessary Wick contractions in Fig.\ \ref{fig:Wick_contractions}. Since our computations are currently based on point-to-all propagators, the resulting matrix is restricted to elements with a local operator at the source. Consequently, $C_{jk}(t)$ are not a square matrices.

To extract the low-lying energy eigenvalues, we performed simultaneous multi-exponential fits to all matrix elements using a truncated spectral decomposition of the correlation matrix,
\begin{equation}
	\label{eq:multiexp} C_{j k}(t) \approx \sum_{n=0}^{N-1} Z_j^n (Z_k^n)^\ast \textrm{e}^{-{E_n} t} .
\end{equation}
$E_n$ denote the energy eigenvalues and $ Z_j^n = \langle \Omega |\op_j | n \rangle $ the overlaps of the corresponding energy eigenstates and the trial states.

%
\subsection{Hadronically Stable $\bar{b}\bar{b}ud$ Tetraquark with $I(J^P)=0(1^+)$}
\begin{figure}[b]
	\centering
	\includegraphics[width=0.6 \textwidth, trim =5 5 5 5 ,clip]{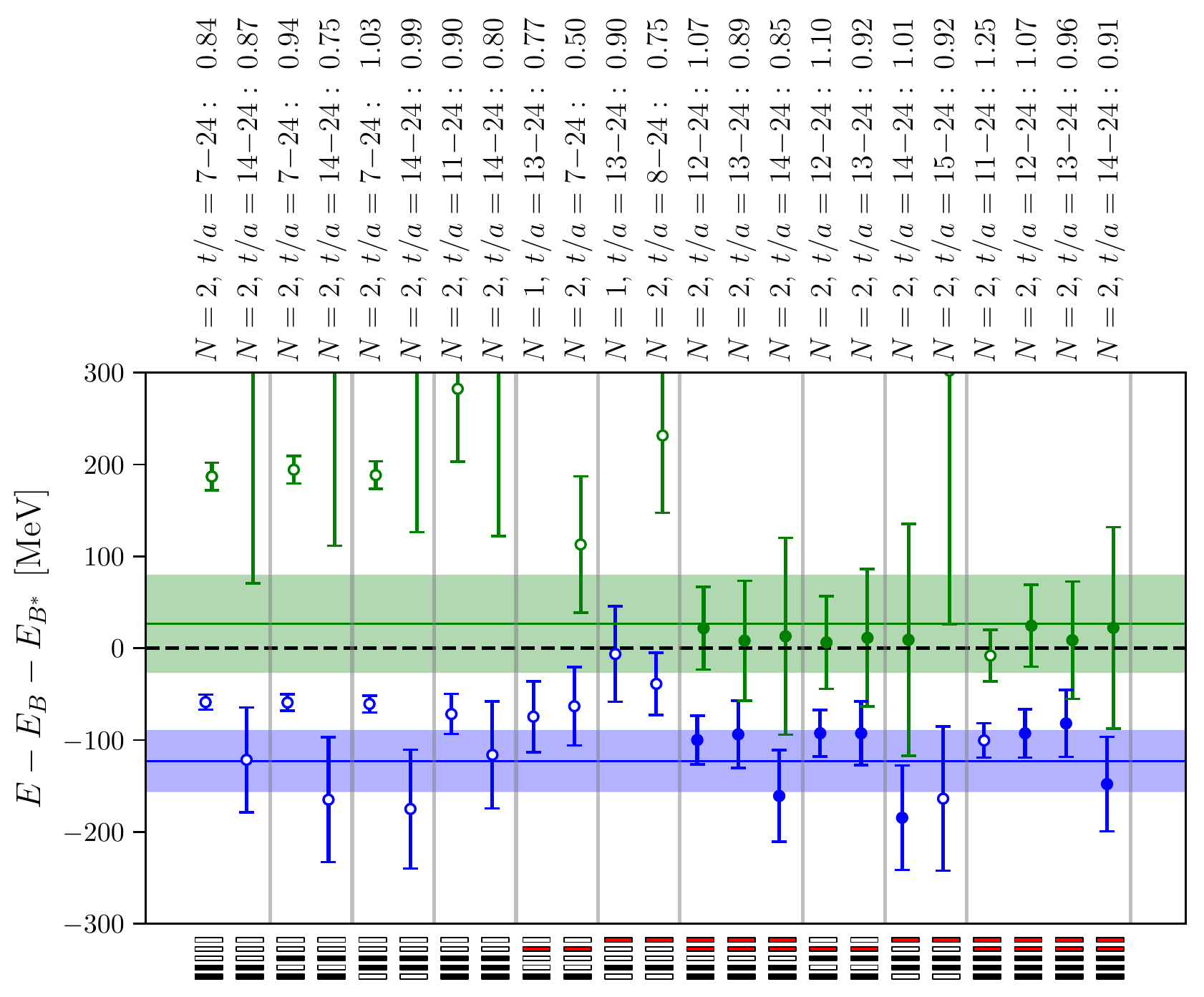}
	\caption{\label{fig:bbud_results}Results for the two lowest energy levels of the $\bar{b}\bar{b}ud$ system with quantum numbers $I(J^P)=0(1^+)$ relative to the $B B^\ast$ threshold (ensemble C005).}
\end{figure}

In Fig.\ \ref{fig:bbud_results} we present fit results for the two lowest energy levels of the $\bar{b}\bar{b}ud$ system with quantum numbers $I(J^P)=0(1^+)$.
The boxes at the bottom of the plot below each fit indicate which operators were included in the correlation matrix. A filled black box represents a local operator and a filled red box a scattering operator. For each operator basis we show the fit results for the ground state in blue and for the first excited state in green, where the energy of the lowest threshold, the $B B^\ast$ threshold, is subtracted. Above the fits, we provide the number of exponentials $N$ used in the fit function \eqref{eq:multiexp}, the temporal fit range and the resulting correlated $\chi^2$.

We obtain a ground-state energy significantly below the relevant $B B^\ast$ threshold, if local as well as scattering operators are included in the correlation matrix. Moreover, the energy of the first excited state is consistent with the $B B^\ast$ threshold. This clearly indicates a hadronically stable tetraquark. A careful analysis based on L\"uscher's finite volume method and a chiral extrapolation including all five ensembles from Tab.\ \ref{tab:configurations} results in a binding energy of $(-128 \pm 24) \, \textrm{MeV}$ with an estimated systematic error below $10 \, \textrm{MeV}$. For details we refer to our recent publication \cite{Leskovec:2019ioa}.

\begin{figure}[htb]
	\includegraphics[width=0.45\linewidth, trim = 30 15 40 40 ,clip]{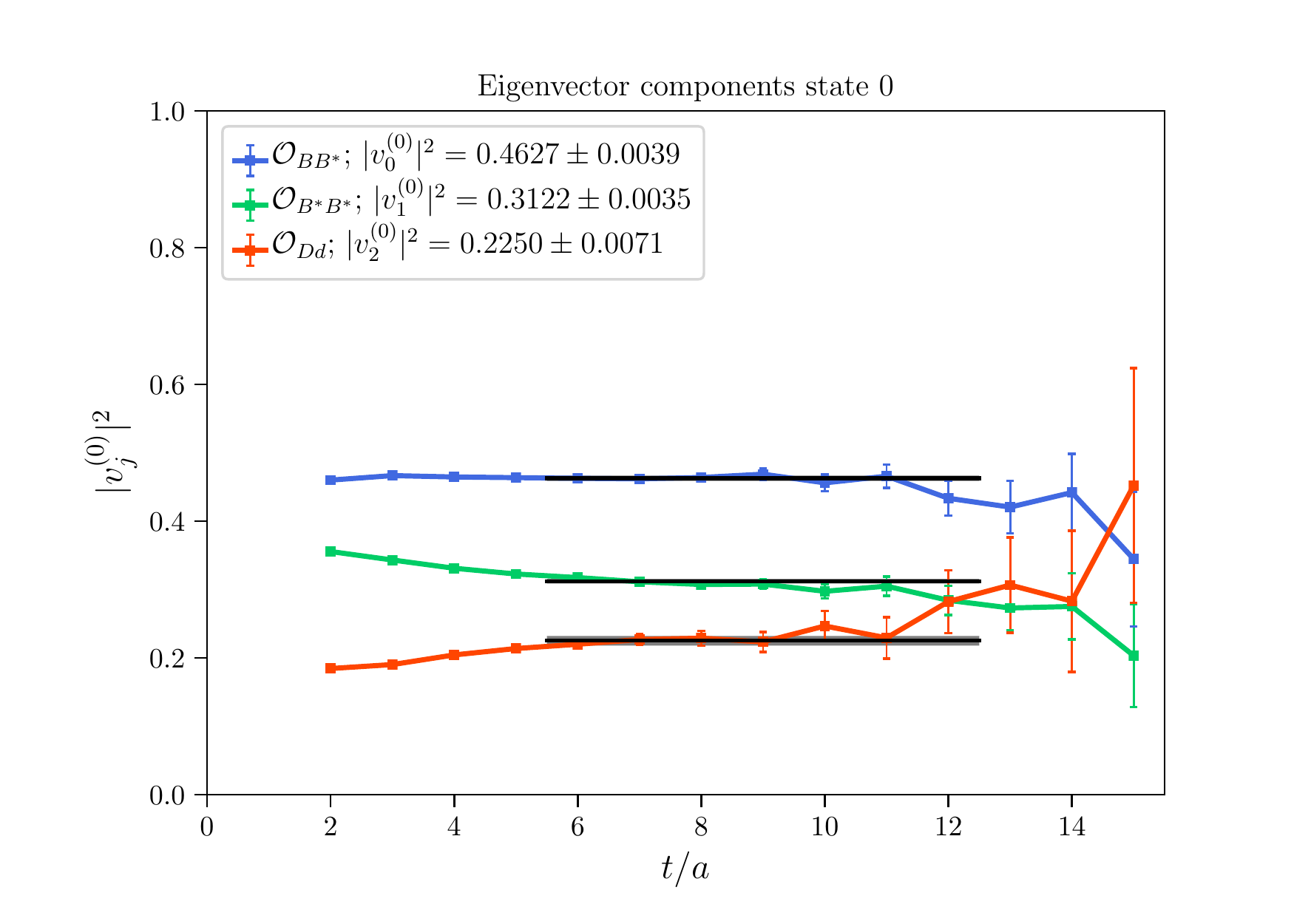}
	\includegraphics[width=0.45\linewidth, trim = 5 20 10 25 ,clip]{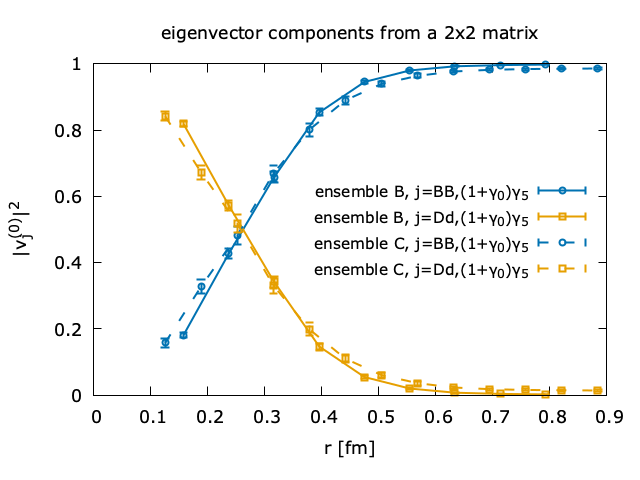}
	\caption{\label{fig:bbud_EV}\textbf{Left:} Normalized eigenvector components of the ground state for the $3 \times 3$ square correlation matrix formed by the local $\bar{b}\bar{b}ud$ operators (ensemble C005). \textbf{Right:} Born-Oppenheimer result  for normalized eigenvector components of a meson-meson and of a diquark-antidiquark interpolating operator as functions of the $\bar b \bar b$ separation $r$ (figure taken from Ref.\ \cite{Bicudo:2021qxj}).}
\end{figure}

We have also solved a standard generalized eigenvalue problem using the $3 \times 3$ square correlation matrix formed by the local operators. In Fig.\ \ref{fig:bbud_EV} (left) we show the normalized eigenvector components of the ground state corresponding to the hadronically stable tetraquark. The plot indicates that the tetraquark is a superposition of $B B$ and $B B^\ast$ meson-meson components and of a diquark-antidiquark component, where the meson-meson contribution dominates with $\sim 77\%$, whereas the diquark-antidiquark contribution is only $\sim 23\%$. It is interesting to compare this result to a recent Born Oppenheimer investigation of the structure of this tetraquark. The main result of Ref.\ \cite{Bicudo:2021qxj}, the eigenvector components of a meson-meson and of a diquark-antidiquark interpolating operator as functions of the $\bar b \bar b$ separation $r$, is shown in Fig.\ \ref{fig:bbud_EV} (right). Multiplying these curves with the radial probability density and integrating over $r$ led to a meson-meson contribution of $\sim 60\%$ and a diquark-antidiquark contribution of $\sim 40 \%$. The results of both approaches agree that both meson-meson and diquark-antidiquark structures are present in the tetraquark with the meson-meson component dominating.

%
\subsection{Hadronically Stable $\bar{b}\bar{b}us$ Tetraquark with $I(J^P)=\tfrac{1}{2}(1^+)$}
\begin{figure}[htb]
	\centering
	\includegraphics[width=0.65 \textwidth, trim = 5 5 5 5 ,clip]{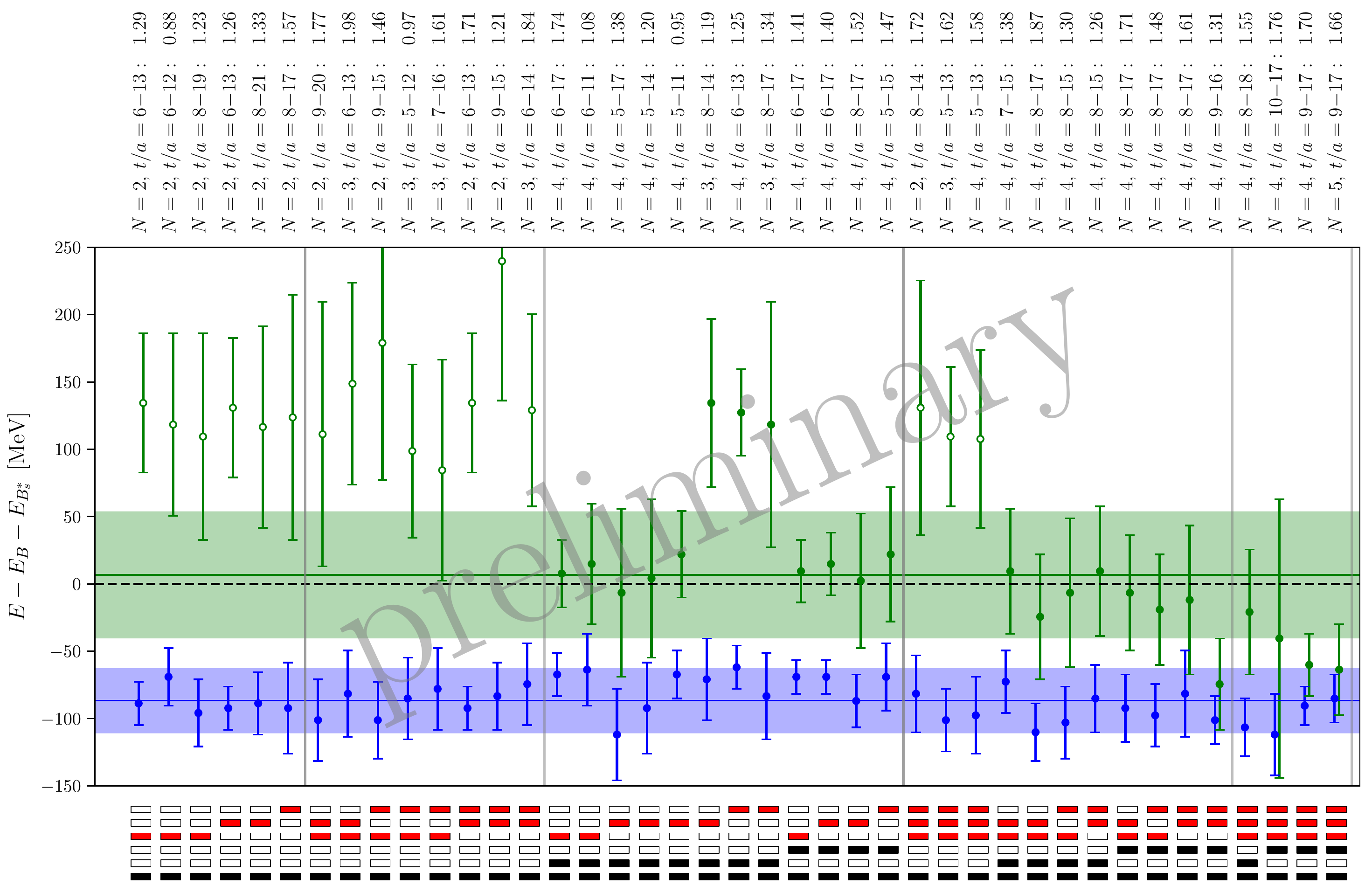}
	\caption{\label{fig:bbus_results}Results for the two lowest energy levels of the $\bar{b}\bar{b}us$ system with quantum numbers $I(J^P)=\tfrac{1}{2}(1^+)$ relative to the $B B_s^\ast$ threshold (ensemble C01).}
\end{figure}

In Fig.\ \ref{fig:bbus_results} we present fit results for the two lowest energy levels of the $\bar{b}\bar{b}us$ system with quantum numbers $I(J^P)=\tfrac{1}{2}(1^+)$. Again there is clear evidence for a hadronically stable tetraquark with a binding energy of $\approx -80 \, \textrm{MeV}$, i.e.\ with a mass clearly below the relevant $B B_s^\ast$ threshold, while the first excited state is consistent with that threshold. This confirms predictions of independent recent lattice-QCD studies using a similar setup \cite{Francis:2016hui,Junnarkar:2018twb}.

\begin{figure}[htb]
	\centering
	\includegraphics[width=0.45\linewidth, page=1, trim = 30 15 40 40 ,clip]{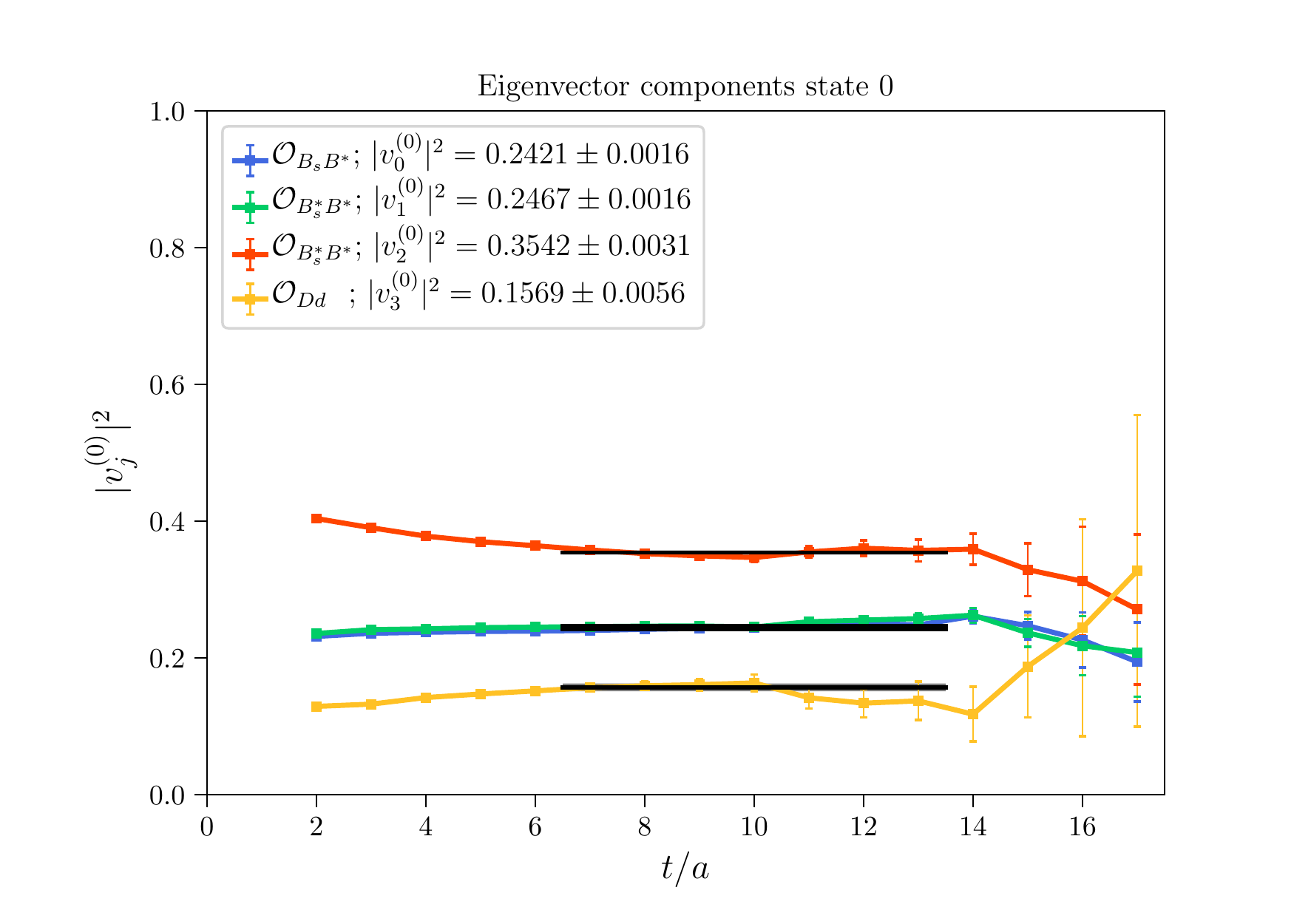}
	\includegraphics[width=0.45\linewidth, page=2, trim = 30 15 40 40 ,clip]{EV_components_bbus.pdf}
	\caption{\label{fig:bbus_EV}Normalized eigenvector components for the $4 \times 4$ square correlation matrix formed by the local $\bar{b}\bar{b}us$ operators (ensemble C01). \textbf{Left:} Ground state. \textbf{Right:} First excitation.}
\end{figure}

As before, we also solved a standard generalized eigenvalue problem using the $4 \times 4$ square correlation matrix formed by the local operators. In Fig.\ \ref{fig:bbus_EV} we show the normalized eigenvector components of the ground state and the first excited state. The left plot indicates that the meson-meson percentage is $\sim 84\%$, i.e.\ somewhat larger than in the $\bar{b}\bar{b}ud$ case, while the diquark-antidiquark percentage is $\sim 16\%$. Interestingly, the $B_s B^\ast$ and $B_s^\ast B$ trial states have almost identical weights and appear either as antisymmetric flavor combination (for the ground state) or as symmetric flavor combination (for the first excitation, when using only local operators). We consider this as strong indication that $\textrm{SU(3)}$ flavor symmetry is approximately fulfilled. This might simplify a scattering analysis, similar to that from Ref.\ \cite{Leskovec:2019ioa}.

%
\subsection{Non-existence of Hadronically Stable $\bar{b}\bar{c}ud$ Tetraquarks with $I(J^P)=0(1^+)$ and $I(J^P)=0(0^+)$}
\begin{figure}[htb]
	\centering
	\includegraphics[width=0.49 \textwidth, trim = 5 5 5 5, clip]{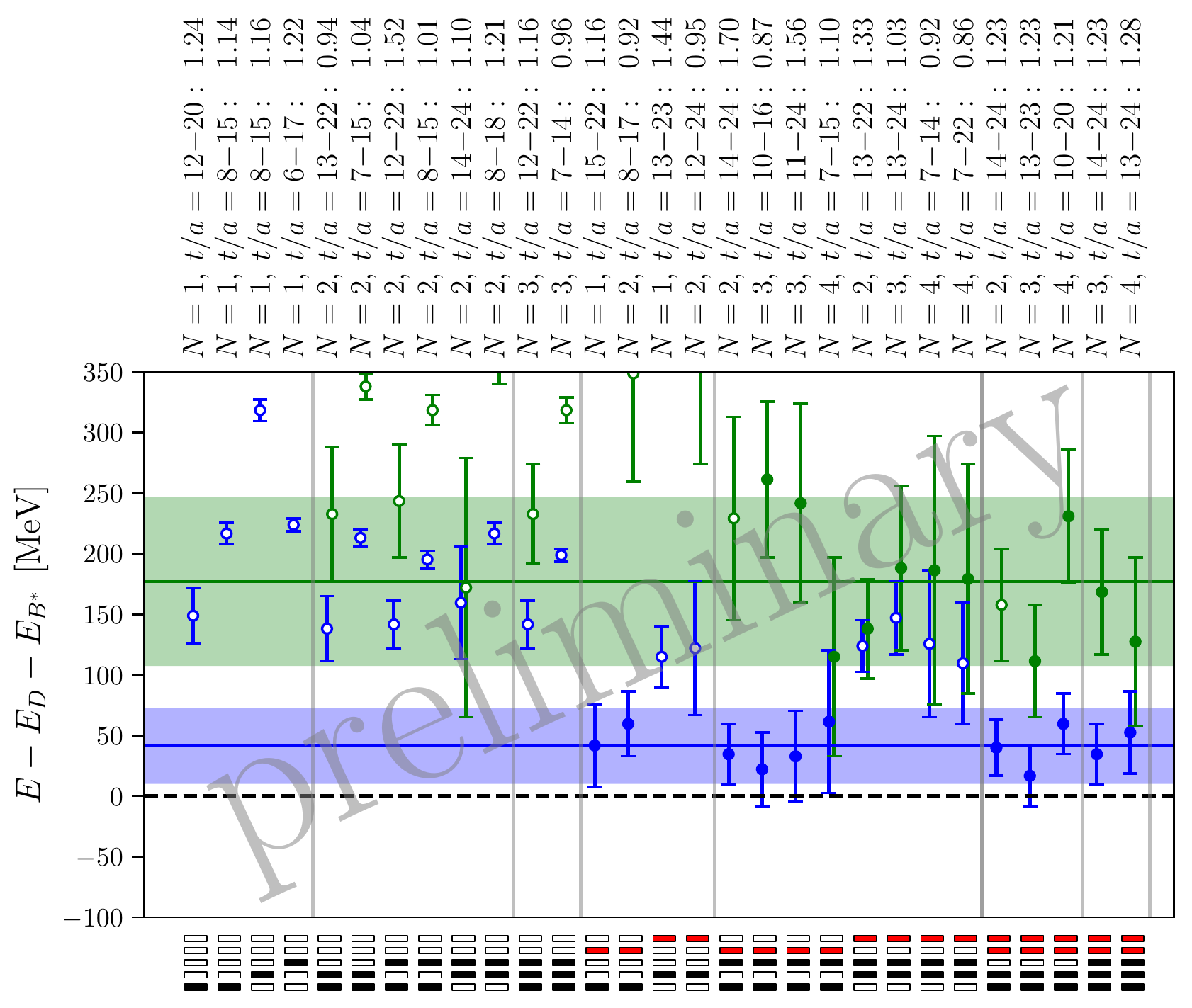}
	\includegraphics[width=0.4 \textwidth, trim = 5 5 5 5  ,clip]{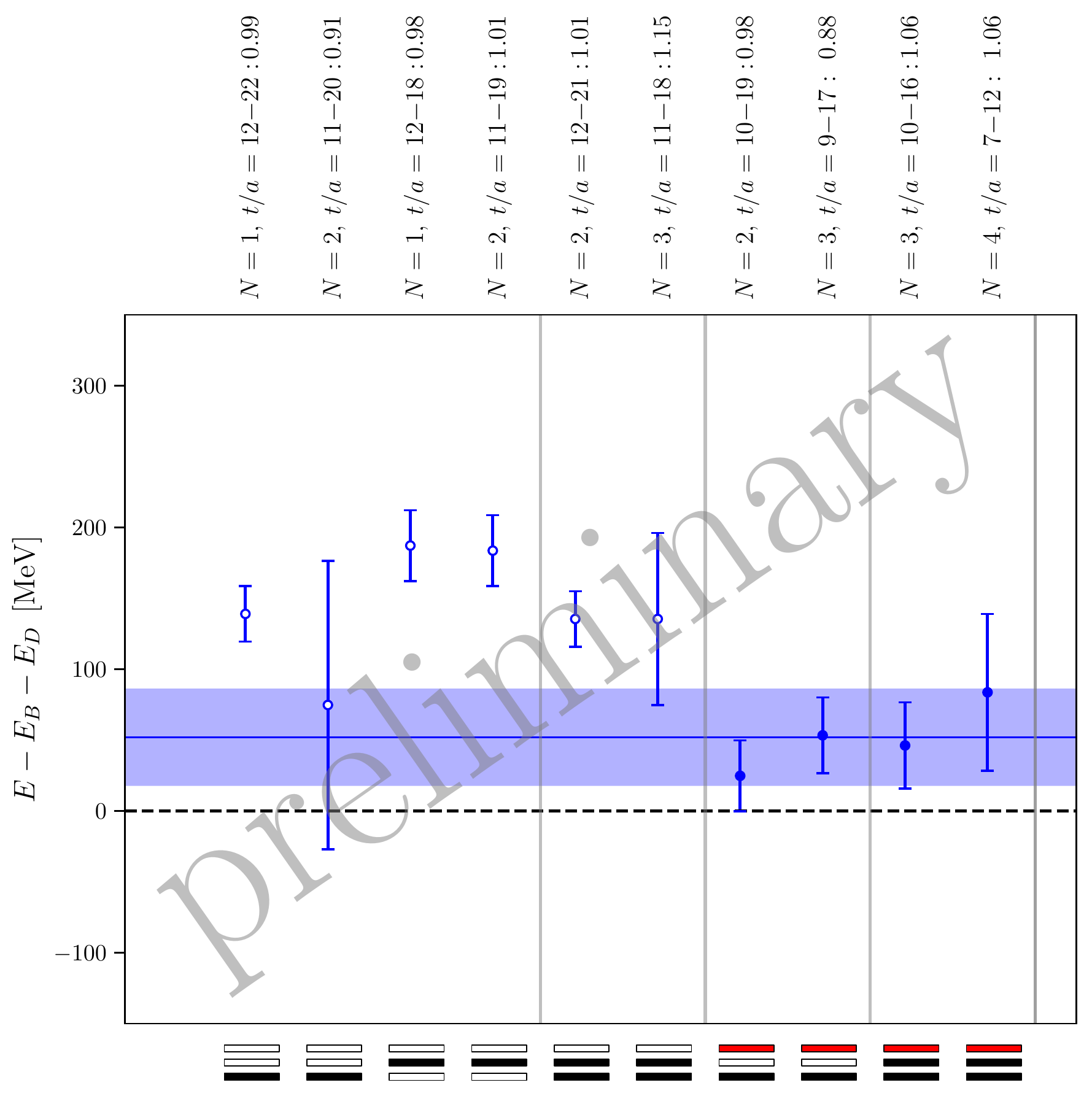}	
	\caption{\label{fig:bcud_results}Results for the lowest energy levels of the $\bar{b}\bar{c}ud$ system. \textbf{Left:} $I(J^P)=0(1^+)$, energy levels relative to the $B^\ast D$ threshold. \textbf{Right:} $I(J^P)=0(0^+)$, energy levels relative to the $B D$ threshold.}
\end{figure}

For quark flavors $\bar{b}\bar{c}ud$ there are two relevant orthogonal channels, either symmetric with respect to the heavy quarks or antisymmetric. A symmetric state corresponds to quantum numbers $I(J^P)=0(1^+)$, an antisymmetric state to $I(J^P)=0(0^+)$. In Fig.\ \ref{fig:bcud_results} we present fit results for both cases: for $I(J^P)=0(1^+)$ the two lowest energy levels, and for $I(J^P)=0(0^+)$ only the ground state energy. In both cases the ground state energy is slightly above, but still consistent with the relevant threshold, i.e.\ there is no indication for a hadronically stable tetraquarks. This supports the findings of Refs.\ \cite{Hudspith:2020tdf}, but contradicts those of Ref.\ \cite{Francis:2018jyb}.

%
\section*{Acknowledgments}

We thank Antje Peters for collaboration in the early stages of this project. We thank the RBC and UKQCD collaborations for providing gauge field ensembles.

L.L.\ acknowledges support from the U.S.\ Department of Energy, Office of Science, through contracts DE-SC0019229 and DE-AC05-06OR23177 (JLAB).
S.M.\ is supported by the U.S.\ Department of Energy, Office of Science, Office of High Energy Physics under Award Number D{E-S}{C0}009913.
M.W.\ acknowledges support by the Heisenberg Programme of the Deutsche Forschungsgemeinschaft (DFG, German Research Foundation) -- project number 399217702.
M.P.\ and M.W.\ acknowledge support by the Deutsche Forschungsgemeinschaft (DFG, German Research Foundation) -- project number 457742095.

 Calculations  on the  GOETHE-HLR  and  on  the  FUCHS-CSC  high-performance  computers  of the Frankfurt University were conducted for this research.  We would like to thank HPC-Hessen, funded by the State Ministry of Higher Education, Research and the Arts, for programming advice.
This research used resources of the National Energy Research Scientific Computing Center (NERSC), a U.S.\ Department of Energy Office of Science User Facility operated under Contract No.\ DE-AC02-05CH11231. This work also used resources at the Texas Advanced Computing Center that are part of the Extreme Science and Engineering Discovery Environment (XSEDE), which is supported by National Science Foundation grant number ACI-1548562.
%
%

\end{document}